\documentclass[reprint,
 superscriptaddress,
 nofootinbib,
 bibnotes,
 amsmath,amssymb, aps,
 prl,
floatfix,
]{revtex4-1}

\usepackage{graphicx}
\usepackage[caption=false]{subfig}
\usepackage{dcolumn}
\usepackage{multirow}
\usepackage{tabularx}
\usepackage{bm}
\usepackage{hyperref}
\usepackage{amsmath}
\usepackage{amssymb}
\usepackage{comment}
\usepackage{color}
\usepackage{enumerate}
\usepackage{booktabs}
\usepackage{siunitx}

\newcommand{\Mpc}{\text{\,Mpc}}
\newcommand{\Gpc}{\text{\,Gpc}}

\newcommand{\eV}{\text{~eV}}

\newcommand{\yr}{\text{~yr}}
\newcommand{\msun}{M_{\odot}}

\def\apj{\ref@jnl{ApJ}}                 

\newcolumntype{L}[1]{>{\hsize=#1\hsize\raggedright\arraybackslash}X}%
\newcolumntype{R}[1]{>{\hsize=#1\hsize\raggedleft\arraybackslash}X}%
\newcolumntype{C}[1]{>{\hsize=#1\hsize\centering\arraybackslash}X}%

\makeatletter
\newcommand{\thickhline}{%
    \noalign {\ifnum 0=`}\fi \hrule height 2pt
    \futurelet \reserved@a \@xhline
}
\newcolumntype{"}{@{\hskip\tabcolsep\vrule width 2pt\hskip\tabcolsep}}
\makeatother

\begin{document}

\title{
Black Hole Mergers and the QCD Axion at Advanced LIGO
}

\author{Asimina Arvanitaki}
\email{aarvanitaki@perimeterinstitute.ca}
\author{Masha Baryakhtar}
\email{mbaryakhtar@perimeterinstitute.ca}
\affiliation{Perimeter Institute for Theoretical Physics, Waterloo, Ontario N2L 2Y5, Canada}
\author{Savas Dimopoulos} 
\email{savas@stanford.edu}
\affiliation{Stanford Institute for Theoretical Physics, Stanford University, Stanford, California 94305, USA}
\author{Sergei Dubovsky}
\email{dubovsky@nyu.edu}
\affiliation{Center for Cosmology and Particle Physics,  New York University New York, NY, 10003, USA}
\author{Robert Lasenby} 
\email{rlasenby@perimeterinstitute.ca}
\affiliation{Perimeter Institute for Theoretical Physics, Waterloo, Ontario N2L 2Y5, Canada}

\begin{abstract}
In the next few years Advanced LIGO (aLIGO) may see gravitational waves
(GWs) from thousands of black hole (BH) mergers. This marks the beginning
of a new precision tool for physics. Here we show how to search for new
physics beyond the standard model using this tool, in particular the QCD
axion in the mass range $\mu_a\sim10^{-14}$ to $10^{-10}\eV$. Axions
(or any bosons) in this mass range cause rapidly rotating BHs to shed
their spin into a large cloud of axions in atomic Bohr orbits around
the BH, through the effect of superradiance (SR). This results in a gap in
the mass vs. spin distribution of BHs when the BH size is comparable to
the axion's Compton wavelength. By measuring the spin and mass of the
merging objects observed at LIGO, we could verify the presence and shape
of the gap in the BH distribution produced by the axion.

The axion cloud can also be discovered through the GWs it radiates
via axion annihilations or level transitions. A blind monochromatic
GW search may reveal up to $10^5$ BHs radiating through axion
annihilations, at distinct frequencies within $\sim3\%$ of $2\mu_a$.
Axion transitions probe heavier axions and may be
observable in future GW observatories. The merger events are perfect
candidates for a targeted GW search. If the final BH has high spin, a SR
cloud may grow and emit monochromatic GWs from axion annihilations. We
may observe the SR evolution in real time.
\end{abstract}

\maketitle

\section{Introduction}
\label{sec:intro}

The LIGO detection of gravitational waves (GWs)~\cite{Abbott:2016blz}
has opened a new window on the universe. In the years to come, GWs
from up to thousands of merger events will reveal a wealth of
information about the hidden lives of black holes and neutron
stars. We also have been given a new precision tool that may diagnose
the presence of new bosonic particles~\cite{Arvanitaki:2009fg}. When
such a particle's Compton wavelength is comparable to the horizon size
of a rotating BH, the superradiance effect~\cite{Zeldovich,Misner:1972kx,Starobinskii} spins down the BH~\cite{Damour:1976kh,Ternov:1978gq,Zouros:1979iw,Detweiler:1980uk},
populating bound Bohr orbits around the BH with an exponentially large
number of particles~\cite{Arvanitaki:2010sy,Arvanitaki:2014wva}.
Astrophysical BHs turn into nature's detectors probing bosons of mass
between $10^{-20}$ and $10^{-10}\eV$. Stellar-mass BHs, such as those
observed by aLIGO, correspond to the upper end of this mass range,
which covers the parameter space for the QCD
axion~\cite{axion1,axion2,axion3} with a decay constant $f_a$ between
the GUT and Planck scales.

The QCD axion was proposed over thirty years ago to explain the
smallness of the neutron electric dipole moment, and has been looked
for ever since. However, SR is not limited to the QCD axion --- it is an excellent probe
of the String Axiverse~\cite{Arvanitaki:2009fg} as well as any other
weakly-interacting boson, such as a dark photon~\cite{Holdom:1985ag,Pani:2012bp},
that lies in the right mass range.
 
In this work we assess how the potentially enormous amount of merger data
collected by aLIGO in the next few years may be used to probe the
effects of SR. A statistical analysis of the spins and masses of merging
BHs can reveal the presence of an axion by the absence of rapidly
rotating BHs. After a merger, the newly-born BH may become a beacon of
monochromatic GW radiation from axion annihilations, providing a
unique opportunity to observe the time evolution of SR. Before we
present the results of our analysis, we review the dynamics of
SR and results from previous work.

\section{Black Hole Superradiance and All-Sky GW Searches}
\label{sec:srstory}

Here we summarize the effects of superradiance on BH evolution (for
detailed discussion, see~\cite{Arvanitaki:2010sy,
  Arvanitaki:2014wva}, as well as~\cite{Brito:2015oca} for a review). We restrict ourselves to the study of
weakly-interacting spin-0 states, with the QCD axion as a primary example.

Axions with Compton wavelength large compared to the size of the BH
have an approximately hydrogenic spectrum of bound states around the
BH with energies
$\omega\approx\mu_a\left(1-\frac{\alpha^2}{2n^2}\right)$, where $\mu_a$ is the axion
mass, $M_{BH}$ the BH mass, and we define $\alpha$ to be the
``fine-structure'' constant of the gravitational ``atom",
\begin{equation}
\alpha\equiv~G_NM_{BH}\mu_a\sim0.22\!\left(\frac{M_{BH}}{30\msun}\right)\!\!\left(\frac{\mu_a}{10^{-12}\eV}\right),
\end{equation}
with $G_N$ Newton's constant~\cite{Detweiler:1980uk,Dolan:2007mj}. Each state is uniquely
characterized by the principal $n$, orbital
$\ell$, and magnetic $m$ quantum numbers.

Such a state is superradiant
(i.e.\ has an occupation number growing with time) if 
\begin{eqnarray}
\frac{\omega}{m}<\Omega_H,
\label{eq:srcondition}
\end{eqnarray}
where $\Omega_H=\frac{1}{2r_g}\frac{a_*}{1+\sqrt{1-a_*^2}}$ is the
angular velocity of the event horizon, $r_g\equiv~G_NM_{BH}$, and
$0\le~a_*<1$ is the dimensionless BH spin.\footnote{The SR condition
  implies $\frac{\alpha}{m}<\frac{1}{2}$, which justifies the
  hydrogenic energy level approximation~\cite{Dolan:2007mj}.} The
simplicity of Eq.~\ref{eq:srcondition} reflects that SR is a
kinematic/thermodynamic phenomenon not unique to
gravity~\cite{Zeldovich,Bekenstein:1998nt}.

When a spinning BH is born, the number of axions in superradiant
levels grows exponentially, seeded by spontaneous emission.  The
growth rate is proportional to the value of the bound-state
wavefunction at the horizon,
$\Gamma_{\mathrm{sr}}\propto\alpha^{4\ell+4}\mu_a$.  The fastest-growing level, generally one
with the minimum $\ell,~m$ such that Eq.~\ref{eq:srcondition} is
satisfied, will extract energy and angular momentum from the BH until
Eq.~\ref{eq:srcondition} is saturated. At that point, the bound state
is occupied by
$N_{\rm{max}}\sim\frac{\Delta a_*}{m}G~M_{BH}^2\sim10^{77}\frac{\Delta a_*}{0.1
  m}\left(\frac{M_{BH}}{10\msun}\right)^2$
axions.  For stellar-mass BHs, e-folding times are as fast as
$\sim100\,\mathrm{sec}$, so energy extraction can occur faster than other
processes such as accretion.  For axion masses much smaller than the
optimum values ($\alpha\ll1$), the growth rate is much slower, while
for much larger masses ($\alpha\gg~1$), satisfying
Eq.~\ref{eq:srcondition} requires $l,m\gg~1$, again giving much slower
growth. Thus, a given BH mass probes a range in mass around
$\mu_a\sim~r_g^{-1}$.

The process repeats for the next-fastest-growing level, until the time
for the next level to grow is longer than the accretion timescale of
the BH or the BH age. Axion self-interactions may modify this picture;
large occupation number in one level may affect the growth of the
others, or lead to axion
emission~\cite{Arvanitaki:2010sy,Arvanitaki:2014wva,Gruzinov:2016hcq}. We
consider masses small enough that, for the QCD axion,
self-interactions are unimportant.

The absence of rapidly rotating old BHs is a signal that SR has taken
place. The spin-mass distribution of BHs should be empty in the
region affected by SR~\cite{Arvanitaki:2009fg,Arvanitaki:2010sy,Arvanitaki:2014wva}.
The handful of high-spin BH measurements in X-ray binaries already
disfavor an axion in the mass range
$6\times10^{-13}\eV~\mathrm{to}~2\times10^{-11}\eV$~\cite{Arvanitaki:2014wva}.

\begin{figure}[t]\vspace{-3cm}
\includegraphics[width = 1 \columnwidth]{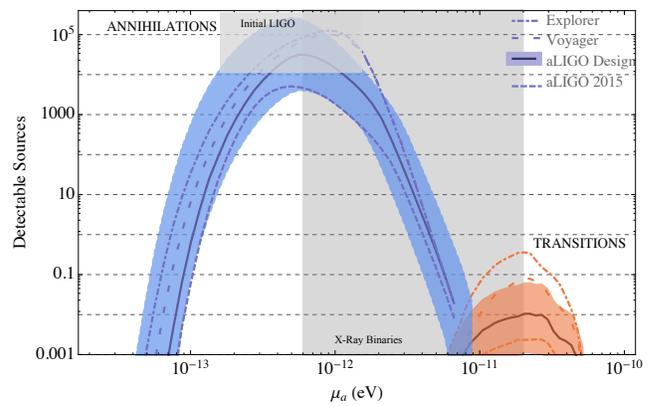}
\vspace{-3cm}
\caption{Expected detectable sources in a blind monochromatic GW
  search, with sensitivity of current aLIGO (dashed), design aLIGO
  (solid), Voyager (wide-dashed) and Cosmic Explorer
  (dot-dashed)~\cite{LIGOWhitePaper} for realistic mass and spin
  distributions, and BH formation rates
  (\cite{Arvanitaki:2014wva}). The shaded bands correspond to the
  range between pessimistic and optimistic BH distributions with
  design aLIGO (distributions as in~\cite{Arvanitaki:2014wva}, with
  the most narrow BH mass distribution removed as it is disfavored by
  the observation of GW150914). The coherent integration time is 2
  days and total time 1 yr.  The annihilation rate has been updated
  using the latest superradiance
  simulations~\cite{Yoshino:2013ofa}. Axion masses in the grayed-out
  region are disfavored by BH spin
  measurements~\cite{Arvanitaki:2014wva}; the most optimistic
  distributions are disfavored by previous null LIGO
  searches~\cite{Aasi:2012fw,Abbott:2016udd,TheLIGOScientific:2016uns}.}\label{fig:blindsearch}
\end{figure}

Axions occupying the bound levels can produce monochromatic GWs in two
ways. Axions can emit a graviton to transition between levels, or two
axions can annihilate into a single
graviton~\cite{Arvanitaki:2009fg,Arvanitaki:2014wva,Okawa:2014nda}.
Annihilations probe axions of mass lighter than $10^{-11}\eV$;
transition signals are largest for axion masses
$\sim10^{-11}-10^{-10}\eV$.  These signals are coherent,
monochromatic, can last 10 years or more, and may be seen in blind
searches for continuous GWs at aLIGO. Fig.~\ref{fig:blindsearch}
summarizes and updates the findings of~\cite{Arvanitaki:2014wva} for
the prospects of those searches. Annihilations provide the most
promising direct probe of SR; assuming exponentially falling BH mass
distributions as in ~\cite{Arvanitaki:2014wva} we expect up to
$\sim10^4$ events at aLIGO coming from annihilations, while axion
transitions become interesting for future detectors\footnote{assuming
  a BH mass distribution falling as a power-law at large mass results
  in an even higher number of annihilation events}.  By updating the
annihilation rates in~\cite{Arvanitaki:2014wva} with the numerical
results of~\cite{Yoshino:2013ofa}, we find that the most optimistic
assumptions about BH mass and spin distributions are already
constrained from null continuous wave searches at initial LIGO~\cite{Aasi:2012fw,Abbott:2016udd,TheLIGOScientific:2016uns}.

In addition to individual monochromatic signals, there would be a
stochastic GW background from unresolved sources. The individual
signals considered in Fig.~\ref{fig:blindsearch} would be concentrated
in a narrow frequency range and stand well above plausible
backgrounds. The stochastic background from axion SR could be
detectable, but as individual signals would likely be seen first we
defer a full discussion to future work.

\section{Statistics of binary BH mergers}
\label{sec:spinstats}

\begin{figure*}[t]
\centering
\includegraphics[width = 1 \columnwidth]{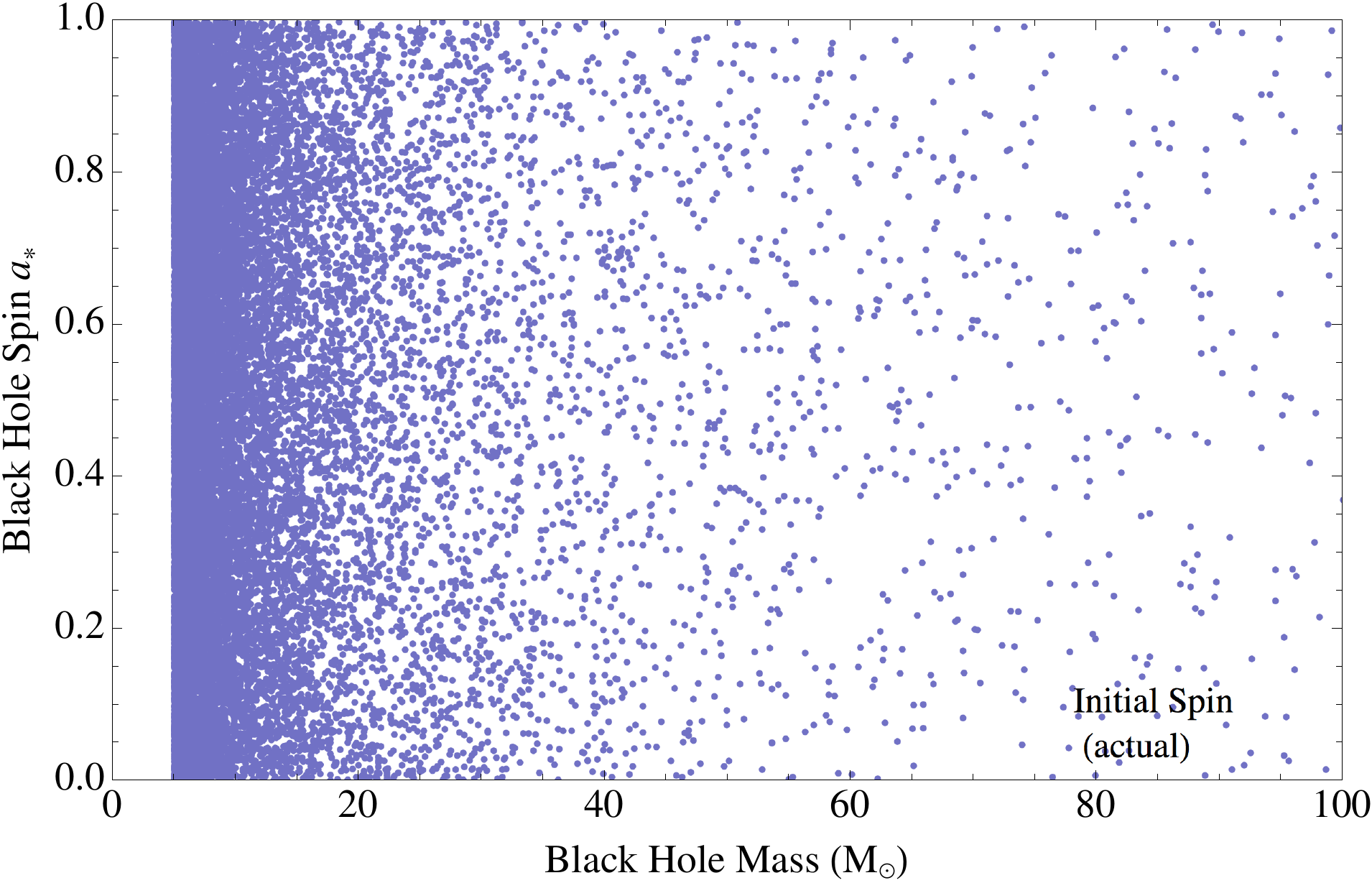}
\centering
\includegraphics[width = 1 \columnwidth]{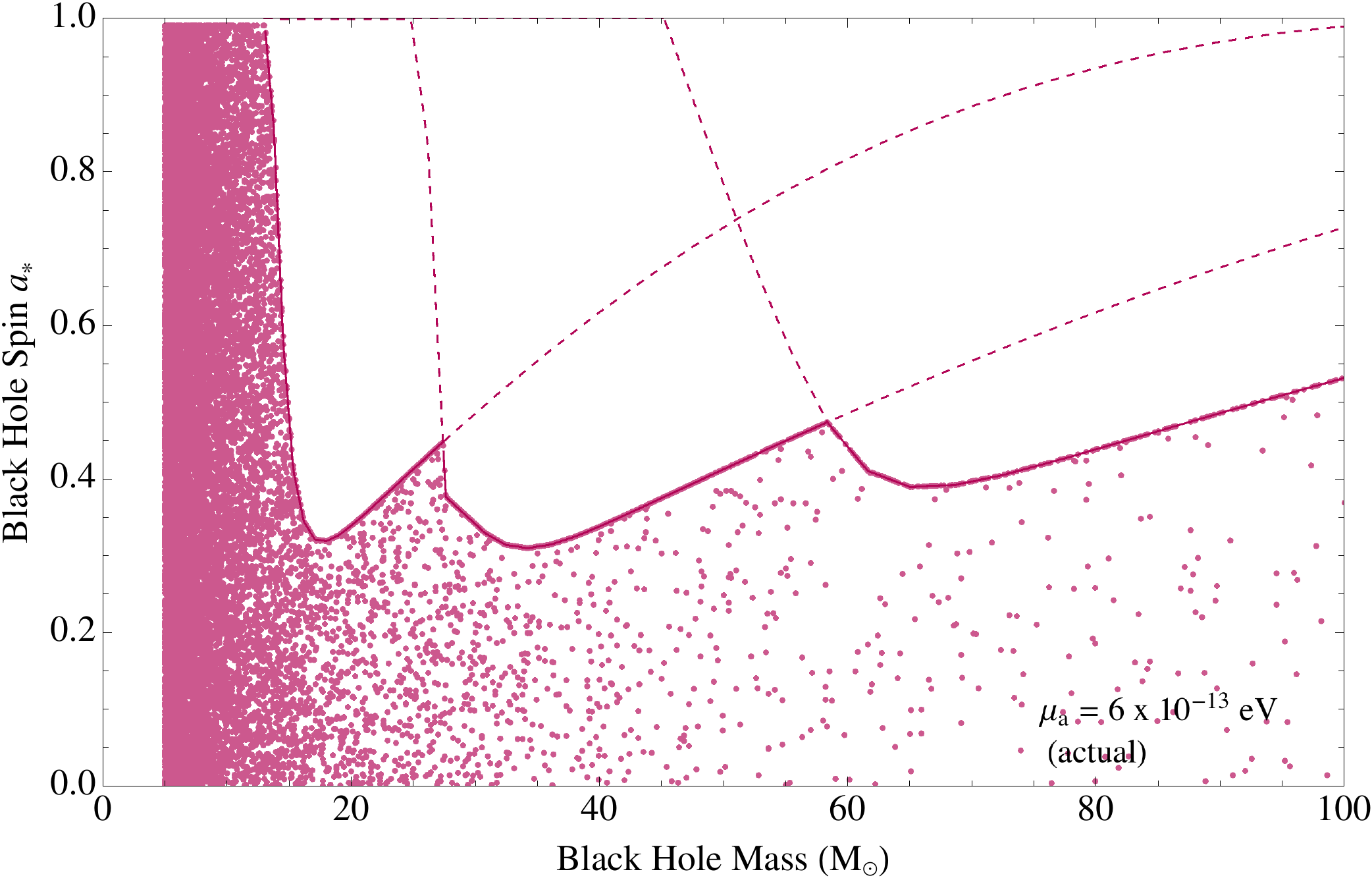}
\centering
\includegraphics[width = 1 \columnwidth]{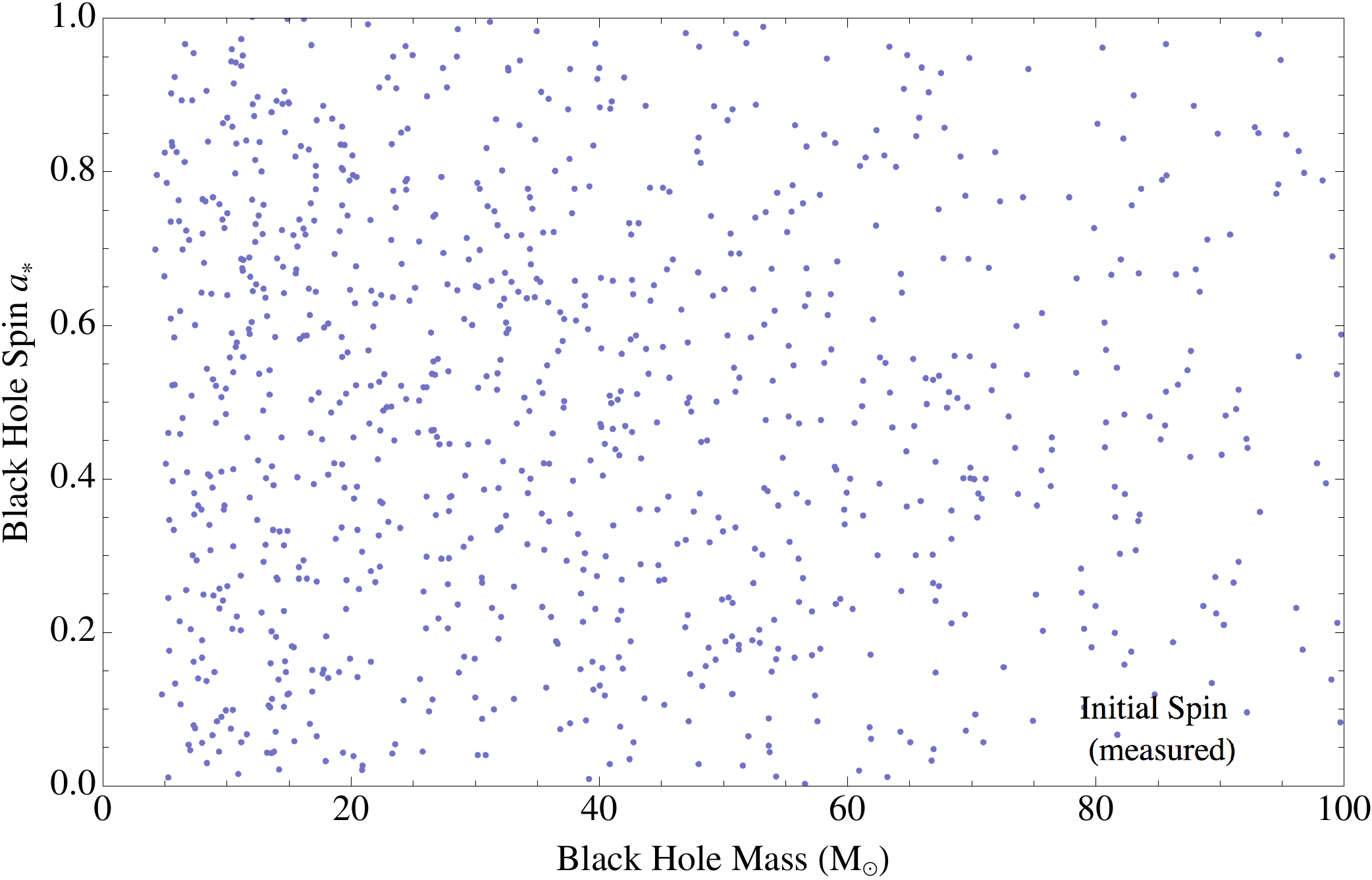}
\centering
\includegraphics[width = 1 \columnwidth]{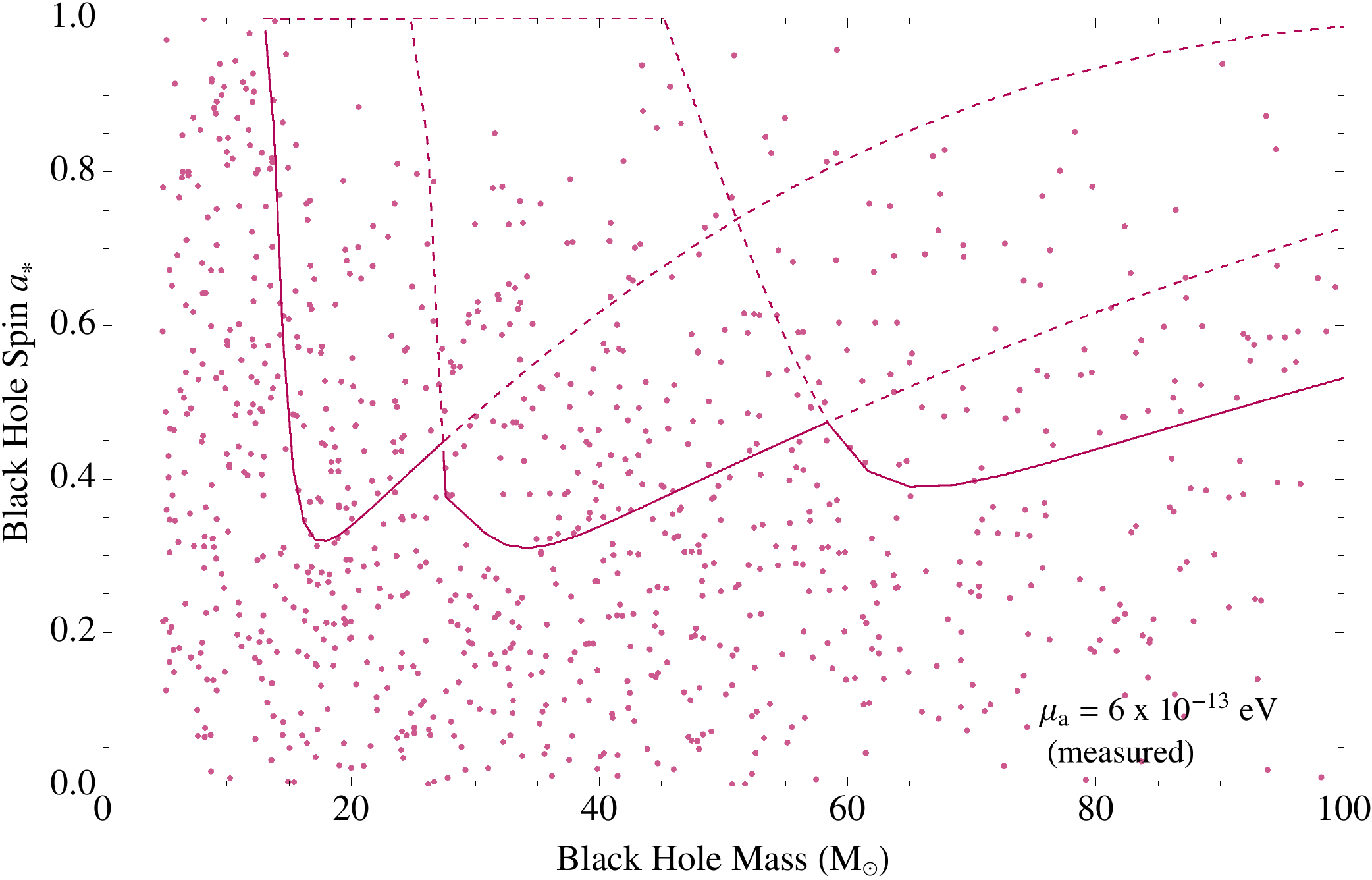}
\caption{Expected distribution of intrinsic (top) and measured
  (bottom) spins and masses of merging BHs in the absence (left) and
  the presence (right) of an axion of mass $6\times10^{-13}$~eV,
  normalized to 1000 events detected at aLIGO. We assume
  $\sigma_M/M\sim10\%$ measurement error in the mass and
  $\sigma_{a_*}\sim0.25$ error in the
  spin~\cite{Vitale:2014mka,*Vitale,Vitale:2016avz}.  We have assumed that all BBHs
  formed at a distance such that they take $10^{10}$ years to
  merge. The theoretical curves shown are boundaries of the regions
  where SR had at most $10^{10}$ years to spin down the BHs, and the
  effect of the companion BH does not significantly affect the SR
  rate.}\label{fig:regge}
\end{figure*}

At design sensitivity, aLIGO is expected to detect 80-1200 binary
black hole (BBH) merger events per year~\cite{Abbott:2016nhf,
  TheLIGOScientific:2016htt,TheLIGOScientific:2016pea}, and measure the masses and spins of the
merging BHs.  A clear signature of superradiance is the absence of
rapidly rotating old BHs in the range influenced by a given axion,
and a large number of BHs populating the curve
$\frac{\omega}{m}=\Omega_H$ for the last level that had time to
grow, as illustrated in Fig.~\ref{fig:regge} (top). We show an
example BH distribution with (right) and without (left) an
axion. Unless otherwise specified, in what follows, we
assume a flat BH spin distribution~\cite{Vitale:2014mka,*Vitale} and a
power-law BH mass distribution
$\rho(M)\propto~M^{-2.35}$~\cite{Abbott:2016nhf} in the absence of an axion,
as were assumed in LIGO analyses.

The histories of BH binaries affect the observed BH distribution in
the mass-spin plane. If the BHs form in an existing binary system and
merge quickly, then superradiant levels might not have had time to
grow to maximal size. In addition, the gravitational perturbation of
one black hole on the other's axion levels can mix superradiating
levels with decaying ones, and may disrupt superradiance
entirely~\cite{Arvanitaki:2010sy,Arvanitaki:2014wva}. On the other
hand, if the binary was formed by
capture~\cite{TheLIGOScientific:2016htt}, the initially isolated BHs
are likely to have had time to superradiate without disruption.

For a given merger time, the BHs are spun down if SR is fast enough to
fully populate the levels before the merger, and the gravitational
perturbation is small enough such that the level-mixing effect on SR
is negligible. Assuming equal mass BHs and initial separation giving
$\tau_{\mathrm{binary}}$ time until the merger (assuming energy loss
through GW emission only), the latter condition for the $\ell=m=1$
level is \cite{ Arvanitaki:2014wva},
\begin{equation}
\alpha\gtrsim0.06\left(\frac{M_{BH}}{30\msun}\right)^{1/15}\left(\frac{10^{10}\yr}{\tau_{\mathrm{binary}}}\right)^{1/15}.
\label{eq:BBHrange-mixing}
\end{equation}
In Fig.~\ref{fig:regge} level mixing is the limiting factor for the
regions affected by $\ell=1,2$ levels, while $\ell=3$ is limited by
the level growth being slower than the binary merger
time.\footnote{The axion cloud is generally destroyed by annihilations
  or falling into the BH without spinning up the BH. Thus, the SR
  saturation lines are a good approximation to the BH's final spin.}

For Fig.~\ref{fig:regge}, we have assumed that the BHs are formed in a
binary, and take $10^{10}$ years to merge. This corresponds to
the largest separation possible, and is the most optimistic scenario
for spin-down, illustrating how strong a signal
could be.  The top panels of Fig.~\ref{fig:regge} present a sample
spin-mass distribution of BBHs with and without an axion.  In the
bottom panels we present the corresponding distributions as seen by
aLIGO, accounting for design detector sensitivity as a function of
total merger mass~\cite{Abbott:1602.03846} and mass and spin
measurement uncertainties~\cite{Vitale:2014mka, Vitale:2016avz}. The large number of
events shown make the lack of rapidly spinning BHs clear.\footnote{It
  may be possible to obtain better spin measurements for BHs in
  BH-NS mergers~\cite{Vitale:2014mka, Vitale:2016avz}, but such events
  have not yet been observed. }

\begin{figure}[t]
	\includegraphics[width = 0.99 \columnwidth]{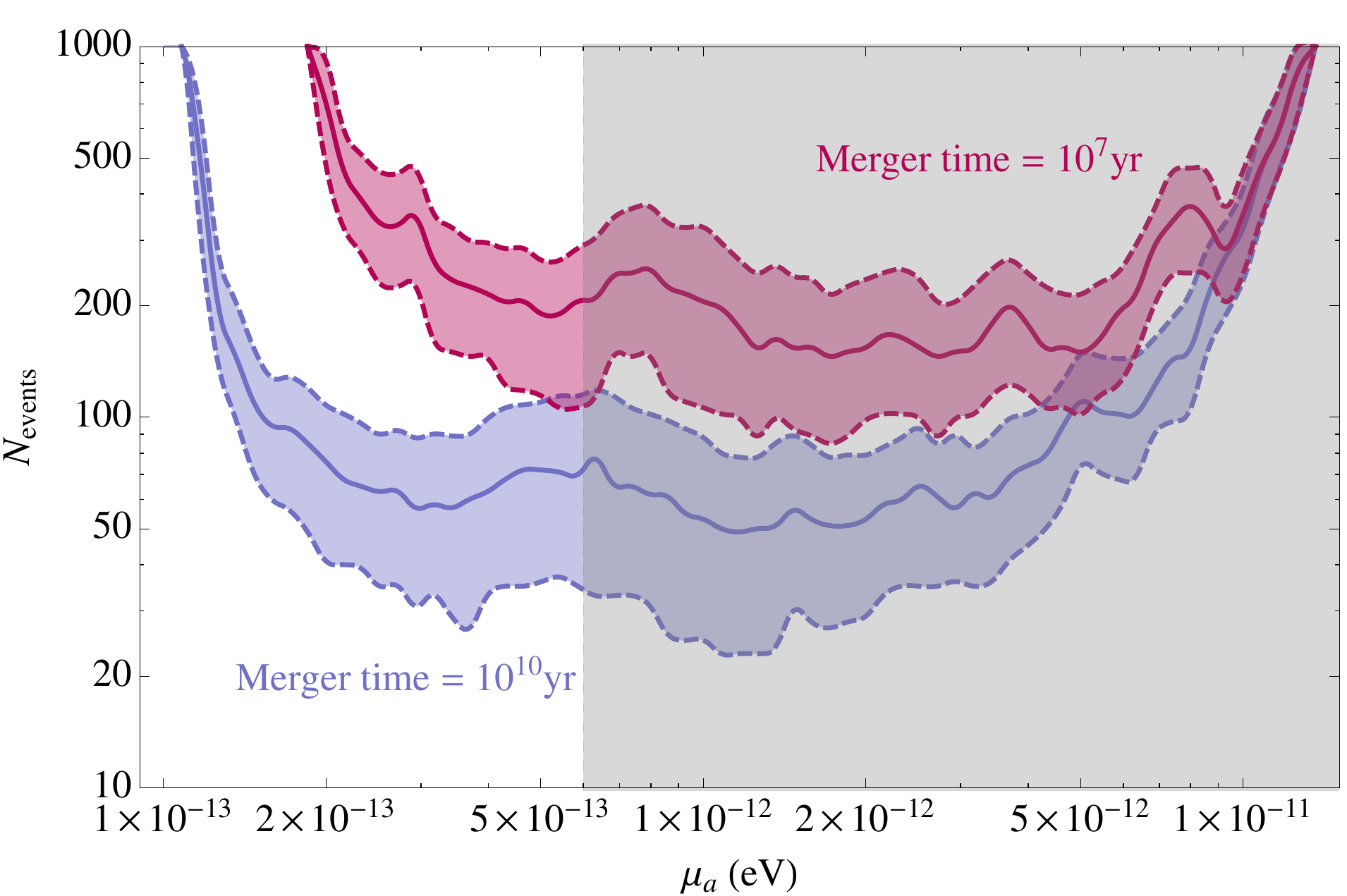}
	\caption{Number of observed events required to show that the
          BH spin distribution varies with BH mass, assuming the
          presence of an axion of mass $\mu_a$.
		  Spin measurement errors of $\sigma_{a_*} = 0.25$
		  are assumed. Blue (red) curves
          correspond to BHs taking $10^{10}\mathrm{yrs}$ ($10^7\mathrm{yrs}$) from
          formation to merger. The solid curves shows the median
          number of events required to reject the
          separable-distribution hypothesis at $2\sigma$.  The
          upper/lower dashed curves show the upper/lower quartiles,
          respectively. The test statistic used is the
          Kolmogorov-Smirnov distance between the spin distributions
          outside and inside a given BH mass range, maximized over
          choice of mass range.  Shaded region is as in
          Fig.~\ref{fig:blindsearch}.  }\label{fig:statistics}
\end{figure}

Even with a relatively small number of events, it may be possible to
infer that the mass-spin distribution has superradiance-like
properties --- for example, that the spin distribution varies with
mass. Fig.~\ref{fig:statistics} shows the number of events at aLIGO
needed to obtain $2\sigma$ evidence for such variation, under the
assumptions explained in the caption. For axion masses between
$\sim 2\times10^{-13}$ eV and $5\times10^{-12}$ eV, we find that good
evidence for a non-separable mass-spin distribution may be obtained
after observing $\mathcal{O}(50)$ events, probing axion masses below
the X-ray binary bounds.

Different assumptions can change the required number of events by
factors of a few.  As shown in Fig.~\ref{fig:statistics}, reducing the
assumed merger time from $10^{10}\mathrm{yrs}$ to $10^{7}\mathrm{yrs}$
(the range suggested by BBH formation models~\cite{Abbott:1602.03846})
increases the number of events necessary and decreases the range of
axion masses probed. A pessimistic assumption of
$\sigma_{a*}~\sim~0.5$ requires $\sim 3-5$ times as many events. Our
error estimates are based on studies of intermediate mass BBHs; at
design-sensitivity LIGO/Virgo detectors, one expects to obtain a 90\%
confidence interval of width $|\Delta a_*| < 0.8$ for total masses up to ~
$600 \msun$, \footnote{for the most pessimistic case of equal BH masses
  and misaligned spins; even better measurements are possible for
  dissimilar masses or aligned spins} and a ~10\% error in mass
determination for an order one fraction of primary black holes masses
\cite{Vitale:2016avz}. These estimates indicate a plausible range of
variation --- a comprehensive analysis, taking into account detailed
mass and spin dependent measurement errors, would require full
simulations.

Of course, dependence of the BH spin distribution on mass may come
from astrophysical effects; if features are seen, more events would be
required to trace out the superradiance contours with accuracy and
determine an axion mass. Third generation observatories can achieve
much higher spin measurement precision ($90\%$ interval of
$|\Delta a_*| < 0.1$ for a majority of events \cite{Vitale:2016icu})
and confirm any features indicated by Advanced LIGO. In addition, if
no features in the mass-spin distribution are seen, we cannot
immediately exclude the presence of an axion, since it may be that
most formation histories did not allow for SR. Nevertheless, a
statistical signal, especially along with other indications of an
axion (e.g. the monochromatic GWs of Fig.~\ref{fig:blindsearch}),
would be suggestive.

\section{Direct Signatures}
\label{sec:directsignatures}

In addition to the wealth of aLIGO measurements of merging black
holes, binary merger events provide a unique opportunity to
observe the birth of a BH.  This BH is the ideal point-source
candidate to observe the evolution of the superradiant instability in
real time.

For transitions, the levels responsible for an appreciable signal take
over a thousand years to grow to large occupation numbers, so are
uninteresting for a followup search. Axion annihilations are the most
promising source of continuous GWs for targeted searches at aLIGO,
with the first level taking from less than a month to up to 10 years
to grow to maximum occupation number. Using the leading-order formula
for the $2$-axion to graviton annihilation rate
$\Gamma_{\rm ann}$ from~\cite{Brito:2014wla}
(see~\cite{Yoshino:2013ofa} for numerical results), the peak GW strain
at Earth from axion annihilations at distance $d$ is~\cite{Arvanitaki:2014wva}
\begin{align}
	\!\! h_{\rm ann}&=\sqrt{\frac{4G_N \Gamma_{\rm ann}N_{\rm max}^2}{2\omega_a
            d^2}} \nonumber\\ &\approx 6\times 10^{-23}\left(\frac{\alpha}{0.3}\right)^7\left (\frac{a_*}{0.9} \right) 
	\!\!  \left(\frac{M_{BH}}{60\msun}\right) \!\!\left(\frac{1 \Mpc}{d}\right),
\end{align}
and lasts for
\begin{equation}
	\tau_{\rm ann}\sim(\Gamma_{\rm ann}N_{\rm max})^{-1}\approx 0.1 \yr \left(\frac{0.3}{\alpha}\right)^{15}\left(\frac{0.9}{a_*}\right)  \!\!  \left(\frac{M_{BH}}{60\msun}\right).
\end{equation}

Correlating these continuous wave emission properties with the spin
and mass of the new BH will be a cross-check on SR predictions.

The reach of aLIGO to an optimal annihilation signal can be as large
as $500\Mpc$ for an axion of mass $10^{-13}\eV$. The reach of aLIGO at
design sensitivity for a typical event is close to $30\Mpc$. In
particular, the final BH of GW150914 with spin of $\sim 0.7$ would
have had to be within $10$~Mpc in order for axion annihilations to be
observable.

\begin{figure}[t]
	\includegraphics[width = 0.99 \columnwidth]{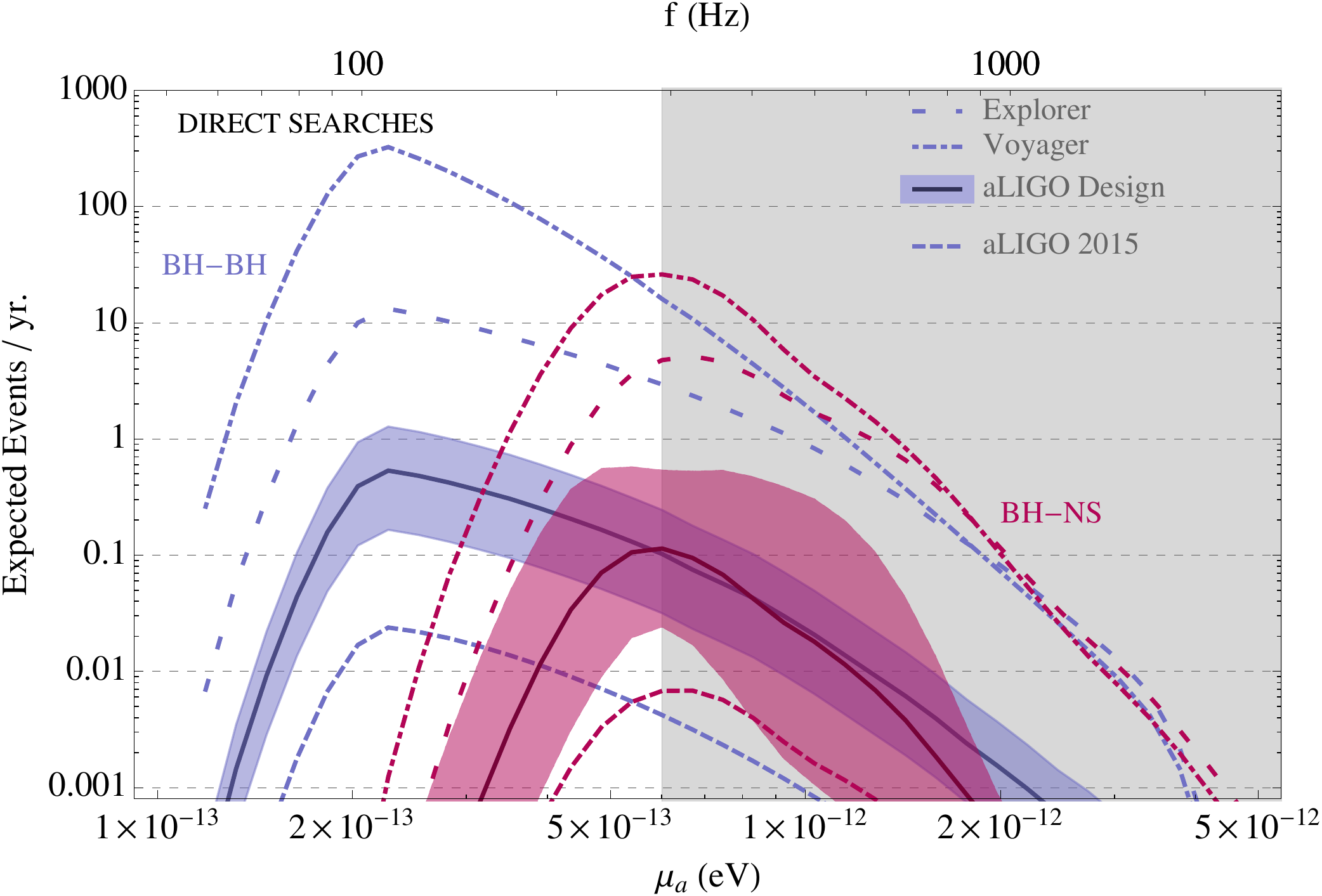}
	\caption{Expected annual annihilation events for aLIGO and
          future observatories from products of BH-NS mergers
          (magenta) or BBH mergers of equal mass (blue).  We assume
          the binary formation mechanism does not allow for
          superradiance. We take $a_*=0$, $M=1.4\msun$ for the NS and
          a power-law mass distribution and flat spin distribution of
          the merging BHs. The bands represent the merger rate
          uncertainty given the observed
          BBHs~\cite{Abbott:2016nhf,TheLIGOScientific:2016pea} and simulations for
          BH-NS (V4l\&V2l in~\cite{Belczynski:2015tba}). We assume a
          coherent integration time of 10 days for BBH and 1 year (or
          up to the duration of the signal) for BH-NS. Shaded region
          is as in Fig.~\ref{fig:blindsearch}.}\label{fig:events}
\end{figure}

In Fig.~\ref{fig:events}, we estimate the number of BBH merger
products emitting observable monochromatic GWs per year, as a function
of the axion mass. The expected number of events is very sensitive to
the spin and mass of the final BH; a linearly-increasing BH spin
distribution increases the expected event rates by a factor of
$\sim 2$ over a flat spin distribution. We estimate the spin of the final BH
with~\cite{Buonanno:2007sv}, assuming equal, aligned initial spins and
equal masses. If SR spun down the initial BHs before the merger, the
final BH will generally not spin quickly enough for SR to produce an
observable signal; for example, we estimate $10^{-3}\mathrm{events/yr}$. at
$\mu_a=2\times10^{-13}\eV$. Only merging BHs for which SR was
inhibited can give rise to a signal observable at aLIGO with an
appreciable rate, and Fig.~\ref{fig:events} assumes this is the case
for an $\mathcal{O}(1)$ fraction of events. There is therefore 
complementarity between the statistical and direct searches --- either
SR spins down enough of these to give a statistical signal, or an
appreciable fraction of post-merger BHs are spinning fast enough to
give direct signals (assuming enough BHs are born with high spin).

Fig.~\ref{fig:events} also shows our expectations for BH-neutron star
(NS) mergers, which have not been observed but are
expected at aLIGO. For BH-NS (as well as NS-NS mergers) we use
expected event rates from numerical simulations,
$1-100\Gpc^{-3}\yr^{-1}$~\cite{Belczynski:2015tba}. Unlike BBHs, we
expect an electromagnetic counterpart~\cite{Lehner:2014asa}, allowing
excellent sky positioning and extending the achievable coherence time
for the monochromatic GW search to the full observation time.  In
addition, BHs produced during these events are lighter, allowing for
searches for heavier axions.

Taking into account the uncertainty in the merger rates, the
expected number of events ranges from $0.01-1\yr^{-1}$ for axion masses
between $2\times10^{-13}-2\times10^{-12}\eV$ coming from BBHs and
BH-NS binaries.

NS-NS mergers create the lightest BHs. These emit high frequency GWs
at the edge of the aLIGO sensitivity curve, and the expected spin of
the final BH is at most $\sim 0.9$~\cite{Kastaun:2013mv}, leading to
low event rates. At design sensitivity, the number of annihilation
events is at best $\sim10^{-3}\yr^{-1}$ for an axion of mass
$8\times10^{-12}\eV$.

Unlike blind searches for isolated BHs, which are dominated by long
signals from our galaxy, searches for post-merger signals are
dominated by stronger signals at tens of Mpc. Thus, event rates will
increase cubically with future strain sensitivity upgrades
(Fig.~\ref{fig:events}), to as many as hundreds of events per year
with projected Explorer sensitivity~\cite{LIGOWhitePaper}.

\section{Conclusion}
\label{sec:conclusion}

The earliest aLIGO signal for an axion is likely to come from
monochromatic GWs in a full-sky survey (Fig. \ref{fig:blindsearch}):
a blind search for continuous waves can potentially discover up to
$\sim10^4$ distinct sources, all within a $\sim3\%$
frequency range (as derived from the SR condition and bound-state
energies). This would be strong evidence for a light boson of mass
equal to half the observed frequency for annihilation signals.  In
contrast, monochromatic GWs produced by astrophysical objects, such as
NSs, are unlikely to cluster within a few percent of a
characteristic frequency. The presence of a unique frequency is a
telltale sign of a new particle.

The statistical search can also lead to early evidence for an axion at
aLIGO. The strength of the statistical evidence will depend on the
formation history, axion mass, and the precision with which spins can
be determined. With good precision, the experimental curve will
approach the theoretical curve (Fig. \ref{fig:regge}, top right) and
the evidence could be compelling. With less precise spin measurements,
the possibility of a yet-unknown standard model mechanism which
disfavors high-spin BHs in a certain mass range would have to be
investigated.

Since the statistical searches and full-sky continuous wave searches
probe a similar axion mass range (Figs. \ref{fig:blindsearch} and
\ref{fig:statistics}), there is the exciting possibility that these
searches may independently indicate the presence of an axion of the
same mass.

The targeted searches of recently formed BHs would be a way to look at
the development of superradiance in real time. This tremendous
possibility may have to wait until aLIGO upgrades. Future aLIGO
upgrades will also make it more likely to observe signals from axion
transitions, which would probe axion masses above $10^{-11}\eV$.
\section{acknowledgements}
We are grateful to Luis Lehner, Salvatore Vitale, and Keith Riles for many
informative and clarifying discussions.  Research at Perimeter
Institute is supported by the Government of Canada through Industry
Canada and by the Province of Ontario through the Ministry of Economic
Development \& Innovation. This work was partially supported by the
National Science Foundation under grant no.~PHYS-1316699 and by the
NSF CAREER award PHY-1352119.

\bibliography{SR-mergers}

\begin{thebibliography}{39}%
\makeatletter
\providecommand \@ifxundefined [1]{%
 \@ifx{#1\undefined}
}%
\providecommand \@ifnum [1]{%
 \ifnum #1\expandafter \@firstoftwo
 \else \expandafter \@secondoftwo
 \fi
}%
\providecommand \@ifx [1]{%
 \ifx #1\expandafter \@firstoftwo
 \else \expandafter \@secondoftwo
 \fi
}%
\providecommand \natexlab [1]{#1}%
\providecommand \enquote  [1]{``#1''}%
\providecommand \bibnamefont  [1]{#1}%
\providecommand \bibfnamefont [1]{#1}%
\providecommand \citenamefont [1]{#1}%
\providecommand \href@noop [0]{\@secondoftwo}%
\providecommand \href [0]{\begingroup \@sanitize@url \@href}%
\providecommand \@href[1]{\@@startlink{#1}\@@href}%
\providecommand \@@href[1]{\endgroup#1\@@endlink}%
\providecommand \@sanitize@url [0]{\catcode `\\12\catcode `\$12\catcode
  `\&12\catcode `\#12\catcode `\^12\catcode `\_12\catcode `\%12\relax}%
\providecommand \@@startlink[1]{}%
\providecommand \@@endlink[0]{}%
\providecommand \url  [0]{\begingroup\@sanitize@url \@url }%
\providecommand \@url [1]{\endgroup\@href {#1}{\urlprefix }}%
\providecommand \urlprefix  [0]{URL }%
\providecommand \Eprint [0]{\href }%
\providecommand \doibase [0]{http://dx.doi.org/}%
\providecommand \selectlanguage [0]{\@gobble}%
\providecommand \bibinfo  [0]{\@secondoftwo}%
\providecommand \bibfield  [0]{\@secondoftwo}%
\providecommand \translation [1]{[#1]}%
\providecommand \BibitemOpen [0]{}%
\providecommand \bibitemStop [0]{}%
\providecommand \bibitemNoStop [0]{.\EOS\space}%
\providecommand \EOS [0]{\spacefactor3000\relax}%
\providecommand \BibitemShut  [1]{\csname bibitem#1\endcsname}%
\let\auto@bib@innerbib\@empty
\bibitem [{\citenamefont {Abbott}\ \emph
  {et~al.}(2016{\natexlab{a}})\citenamefont {Abbott} \emph
  {et~al.}}]{Abbott:2016blz}%
  \BibitemOpen
  \bibfield  {author} {\bibinfo {author} {\bibfnamefont {B.~P.}\ \bibnamefont
  {Abbott}} \emph {et~al.} (\bibinfo {collaboration} {Virgo, LIGO
  Scientific}),\ }\href {\doibase 10.1103/PhysRevLett.116.061102} {\bibfield
  {journal} {\bibinfo  {journal} {Phys. Rev. Lett.}\ }\textbf {\bibinfo
  {volume} {116}},\ \bibinfo {pages} {061102} (\bibinfo {year}
  {2016}{\natexlab{a}})},\ \Eprint {http://arxiv.org/abs/1602.03837}
  {arXiv:1602.03837 [gr-qc]} \BibitemShut {NoStop}%
\bibitem [{\citenamefont {Arvanitaki}\ \emph {et~al.}(2010)\citenamefont
  {Arvanitaki}, \citenamefont {Dimopoulos}, \citenamefont {Dubovsky},
  \citenamefont {Kaloper},\ and\ \citenamefont
  {March-Russell}}]{Arvanitaki:2009fg}%
  \BibitemOpen
  \bibfield  {author} {\bibinfo {author} {\bibfnamefont {A.}~\bibnamefont
  {Arvanitaki}}, \bibinfo {author} {\bibfnamefont {S.}~\bibnamefont
  {Dimopoulos}}, \bibinfo {author} {\bibfnamefont {S.}~\bibnamefont
  {Dubovsky}}, \bibinfo {author} {\bibfnamefont {N.}~\bibnamefont {Kaloper}}, \
  and\ \bibinfo {author} {\bibfnamefont {J.}~\bibnamefont {March-Russell}},\
  }\href {\doibase 10.1103/PhysRevD.81.123530} {\bibfield  {journal} {\bibinfo
  {journal} {Phys. Rev.}\ }\textbf {\bibinfo {volume} {D81}},\ \bibinfo {pages}
  {123530} (\bibinfo {year} {2010})},\ \Eprint {http://arxiv.org/abs/0905.4720}
  {arXiv:0905.4720 [hep-th]} \BibitemShut {NoStop}%
\bibitem [{\citenamefont {Zeldovich}(1971)}]{Zeldovich}%
  \BibitemOpen
  \bibfield  {author} {\bibinfo {author} {\bibfnamefont {Y.~B.}\ \bibnamefont
  {Zeldovich}},\ }\href@noop {} {\bibfield  {journal} {\bibinfo  {journal}
  {JETP Lett.}\ }\textbf {\bibinfo {volume} {14}},\ \bibinfo {pages} {180}
  (\bibinfo {year} {1971})}\BibitemShut {NoStop}%
\bibitem [{\citenamefont {Misner}(1972)}]{Misner:1972kx}%
  \BibitemOpen
  \bibfield  {author} {\bibinfo {author} {\bibfnamefont {C.~W.}\ \bibnamefont
  {Misner}},\ }\href {\doibase 10.1103/PhysRevLett.28.994} {\bibfield
  {journal} {\bibinfo  {journal} {Phys. Rev. Lett.}\ }\textbf {\bibinfo
  {volume} {28}},\ \bibinfo {pages} {994} (\bibinfo {year} {1972})}\BibitemShut
  {NoStop}%
\bibitem [{\citenamefont {Starobinskii}(1973)}]{Starobinskii}%
  \BibitemOpen
  \bibfield  {author} {\bibinfo {author} {\bibfnamefont {A.}~\bibnamefont
  {Starobinskii}},\ }\href@noop {} {\bibfield  {journal} {\bibinfo  {journal}
  {Soviet Phys. JETP}\ }\textbf {\bibinfo {volume} {37}},\ \bibinfo {pages}
  {28} (\bibinfo {year} {1973})}\BibitemShut {NoStop}%
\bibitem [{\citenamefont {Damour}\ \emph {et~al.}(1976)\citenamefont {Damour},
  \citenamefont {Deruelle},\ and\ \citenamefont {Ruffini}}]{Damour:1976kh}%
  \BibitemOpen
  \bibfield  {author} {\bibinfo {author} {\bibfnamefont {T.}~\bibnamefont
  {Damour}}, \bibinfo {author} {\bibfnamefont {N.}~\bibnamefont {Deruelle}}, \
  and\ \bibinfo {author} {\bibfnamefont {R.}~\bibnamefont {Ruffini}},\ }\href
  {\doibase 10.1007/BF02725534} {\bibfield  {journal} {\bibinfo  {journal}
  {Lett. Nuovo Cim.}\ }\textbf {\bibinfo {volume} {15}},\ \bibinfo {pages}
  {257} (\bibinfo {year} {1976})}\BibitemShut {NoStop}%
\bibitem [{\citenamefont {Ternov}\ \emph {et~al.}(1978)\citenamefont {Ternov},
  \citenamefont {Khalilov}, \citenamefont {Chizhov},\ and\ \citenamefont
  {Gaina}}]{Ternov:1978gq}%
  \BibitemOpen
  \bibfield  {author} {\bibinfo {author} {\bibfnamefont {I.~M.}\ \bibnamefont
  {Ternov}}, \bibinfo {author} {\bibfnamefont {V.~R.}\ \bibnamefont
  {Khalilov}}, \bibinfo {author} {\bibfnamefont {G.~A.}\ \bibnamefont
  {Chizhov}}, \ and\ \bibinfo {author} {\bibfnamefont {A.~B.}\ \bibnamefont
  {Gaina}},\ }\href {\doibase 10.1007/BF00894575} {\bibfield  {journal}
  {\bibinfo  {journal} {Sov. Phys. J.}\ }\textbf {\bibinfo {volume} {21}},\
  \bibinfo {pages} {1200} (\bibinfo {year} {1978})},\ \bibinfo {note} {[Izv.
  Vuz. Fiz.21N9,109(1978)]}\BibitemShut {NoStop}%
\bibitem [{\citenamefont {Zouros}\ and\ \citenamefont
  {Eardley}(1979)}]{Zouros:1979iw}%
  \BibitemOpen
  \bibfield  {author} {\bibinfo {author} {\bibfnamefont {T.~J.~M.}\
  \bibnamefont {Zouros}}\ and\ \bibinfo {author} {\bibfnamefont {D.~M.}\
  \bibnamefont {Eardley}},\ }\href {\doibase 10.1016/0003-4916(79)90237-9}
  {\bibfield  {journal} {\bibinfo  {journal} {Annals Phys.}\ }\textbf {\bibinfo
  {volume} {118}},\ \bibinfo {pages} {139} (\bibinfo {year}
  {1979})}\BibitemShut {NoStop}%
\bibitem [{\citenamefont {Detweiler}(1980)}]{Detweiler:1980uk}%
  \BibitemOpen
  \bibfield  {author} {\bibinfo {author} {\bibfnamefont {S.~L.}\ \bibnamefont
  {Detweiler}},\ }\href {\doibase 10.1103/PhysRevD.22.2323} {\bibfield
  {journal} {\bibinfo  {journal} {Phys. Rev.}\ }\textbf {\bibinfo {volume}
  {D22}},\ \bibinfo {pages} {2323} (\bibinfo {year} {1980})}\BibitemShut
  {NoStop}%
\bibitem [{\citenamefont {Arvanitaki}\ and\ \citenamefont
  {Dubovsky}(2011)}]{Arvanitaki:2010sy}%
  \BibitemOpen
  \bibfield  {author} {\bibinfo {author} {\bibfnamefont {A.}~\bibnamefont
  {Arvanitaki}}\ and\ \bibinfo {author} {\bibfnamefont {S.}~\bibnamefont
  {Dubovsky}},\ }\href {\doibase 10.1103/PhysRevD.83.044026} {\bibfield
  {journal} {\bibinfo  {journal} {Phys. Rev.}\ }\textbf {\bibinfo {volume}
  {D83}},\ \bibinfo {pages} {044026} (\bibinfo {year} {2011})},\ \Eprint
  {http://arxiv.org/abs/1004.3558} {arXiv:1004.3558 [hep-th]} \BibitemShut
  {NoStop}%
\bibitem [{\citenamefont {Arvanitaki}\ \emph {et~al.}(2015)\citenamefont
  {Arvanitaki}, \citenamefont {Baryakhtar},\ and\ \citenamefont
  {Huang}}]{Arvanitaki:2014wva}%
  \BibitemOpen
  \bibfield  {author} {\bibinfo {author} {\bibfnamefont {A.}~\bibnamefont
  {Arvanitaki}}, \bibinfo {author} {\bibfnamefont {M.}~\bibnamefont
  {Baryakhtar}}, \ and\ \bibinfo {author} {\bibfnamefont {X.}~\bibnamefont
  {Huang}},\ }\href {\doibase 10.1103/PhysRevD.91.084011} {\bibfield  {journal}
  {\bibinfo  {journal} {Phys. Rev.}\ }\textbf {\bibinfo {volume} {D91}},\
  \bibinfo {pages} {084011} (\bibinfo {year} {2015})},\ \Eprint
  {http://arxiv.org/abs/1411.2263} {arXiv:1411.2263 [hep-ph]} \BibitemShut
  {NoStop}%
\bibitem [{\citenamefont {Weinberg}(1978)}]{axion1}%
  \BibitemOpen
  \bibfield  {author} {\bibinfo {author} {\bibfnamefont {S.}~\bibnamefont
  {Weinberg}},\ }\href {\doibase 10.1103/PhysRevLett.40.223} {\bibfield
  {journal} {\bibinfo  {journal} {Phys.Rev.Lett.}\ }\textbf {\bibinfo {volume}
  {40}},\ \bibinfo {pages} {223} (\bibinfo {year} {1978})}\BibitemShut
  {NoStop}%
\bibitem [{\citenamefont {Wilczek}(1978)}]{axion2}%
  \BibitemOpen
  \bibfield  {author} {\bibinfo {author} {\bibfnamefont {F.}~\bibnamefont
  {Wilczek}},\ }\href {\doibase 10.1103/PhysRevLett.40.279} {\bibfield
  {journal} {\bibinfo  {journal} {Phys.Rev.Lett.}\ }\textbf {\bibinfo {volume}
  {40}},\ \bibinfo {pages} {279} (\bibinfo {year} {1978})}\BibitemShut
  {NoStop}%
\bibitem [{\citenamefont {Peccei}\ and\ \citenamefont {Quinn}(1977)}]{axion3}%
  \BibitemOpen
  \bibfield  {author} {\bibinfo {author} {\bibfnamefont {R.}~\bibnamefont
  {Peccei}}\ and\ \bibinfo {author} {\bibfnamefont {H.~R.}\ \bibnamefont
  {Quinn}},\ }\href {\doibase 10.1103/PhysRevLett.38.1440} {\bibfield
  {journal} {\bibinfo  {journal} {Phys.Rev.Lett.}\ }\textbf {\bibinfo {volume}
  {38}},\ \bibinfo {pages} {1440} (\bibinfo {year} {1977})}\BibitemShut
  {NoStop}%
\bibitem [{\citenamefont {Holdom}(1986)}]{Holdom:1985ag}%
  \BibitemOpen
  \bibfield  {author} {\bibinfo {author} {\bibfnamefont {B.}~\bibnamefont
  {Holdom}},\ }\href {\doibase 10.1016/0370-2693(86)91377-8} {\bibfield
  {journal} {\bibinfo  {journal} {Phys. Lett.}\ }\textbf {\bibinfo {volume}
  {B166}},\ \bibinfo {pages} {196} (\bibinfo {year} {1986})}\BibitemShut
  {NoStop}%
\bibitem [{\citenamefont {Pani}\ \emph {et~al.}(2012)\citenamefont {Pani},
  \citenamefont {Cardoso}, \citenamefont {Gualtieri}, \citenamefont {Berti},\
  and\ \citenamefont {Ishibashi}}]{Pani:2012bp}%
  \BibitemOpen
  \bibfield  {author} {\bibinfo {author} {\bibfnamefont {P.}~\bibnamefont
  {Pani}}, \bibinfo {author} {\bibfnamefont {V.}~\bibnamefont {Cardoso}},
  \bibinfo {author} {\bibfnamefont {L.}~\bibnamefont {Gualtieri}}, \bibinfo
  {author} {\bibfnamefont {E.}~\bibnamefont {Berti}}, \ and\ \bibinfo {author}
  {\bibfnamefont {A.}~\bibnamefont {Ishibashi}},\ }\href {\doibase
  10.1103/PhysRevD.86.104017} {\bibfield  {journal} {\bibinfo  {journal} {Phys.
  Rev.}\ }\textbf {\bibinfo {volume} {D86}},\ \bibinfo {pages} {104017}
  (\bibinfo {year} {2012})},\ \Eprint {http://arxiv.org/abs/1209.0773}
  {arXiv:1209.0773 [gr-qc]} \BibitemShut {NoStop}%
\bibitem [{\citenamefont {Brito}\ \emph
  {et~al.}(2015{\natexlab{a}})\citenamefont {Brito}, \citenamefont {Cardoso},\
  and\ \citenamefont {Pani}}]{Brito:2015oca}%
  \BibitemOpen
  \bibfield  {author} {\bibinfo {author} {\bibfnamefont {R.}~\bibnamefont
  {Brito}}, \bibinfo {author} {\bibfnamefont {V.}~\bibnamefont {Cardoso}}, \
  and\ \bibinfo {author} {\bibfnamefont {P.}~\bibnamefont {Pani}},\ }\href
  {\doibase 10.1007/978-3-319-19000-6} {\bibfield  {journal} {\bibinfo
  {journal} {Lect. Notes Phys.}\ }\textbf {\bibinfo {volume} {906}},\ \bibinfo
  {pages} {pp.1} (\bibinfo {year} {2015}{\natexlab{a}})},\ \Eprint
  {http://arxiv.org/abs/1501.06570} {arXiv:1501.06570 [gr-qc]} \BibitemShut
  {NoStop}%
\bibitem [{\citenamefont {Dolan}(2007)}]{Dolan:2007mj}%
  \BibitemOpen
  \bibfield  {author} {\bibinfo {author} {\bibfnamefont {S.~R.}\ \bibnamefont
  {Dolan}},\ }\href {\doibase 10.1103/PhysRevD.76.084001} {\bibfield  {journal}
  {\bibinfo  {journal} {Phys. Rev.}\ }\textbf {\bibinfo {volume} {D76}},\
  \bibinfo {pages} {084001} (\bibinfo {year} {2007})},\ \Eprint
  {http://arxiv.org/abs/0705.2880} {arXiv:0705.2880 [gr-qc]} \BibitemShut
  {NoStop}%
\bibitem [{\citenamefont {Bekenstein}\ and\ \citenamefont
  {Schiffer}(1998)}]{Bekenstein:1998nt}%
  \BibitemOpen
  \bibfield  {author} {\bibinfo {author} {\bibfnamefont {J.~D.}\ \bibnamefont
  {Bekenstein}}\ and\ \bibinfo {author} {\bibfnamefont {M.}~\bibnamefont
  {Schiffer}},\ }\href {\doibase 10.1103/PhysRevD.58.064014} {\bibfield
  {journal} {\bibinfo  {journal} {Phys. Rev.}\ }\textbf {\bibinfo {volume}
  {D58}},\ \bibinfo {pages} {064014} (\bibinfo {year} {1998})},\ \Eprint
  {http://arxiv.org/abs/gr-qc/9803033} {arXiv:gr-qc/9803033 [gr-qc]}
  \BibitemShut {NoStop}%
\bibitem [{\citenamefont {Gruzinov}(2016)}]{Gruzinov:2016hcq}%
  \BibitemOpen
  \bibfield  {author} {\bibinfo {author} {\bibfnamefont {A.}~\bibnamefont
  {Gruzinov}},\ }\href@noop {} {\  (\bibinfo {year} {2016})},\ \Eprint
  {http://arxiv.org/abs/1604.06422} {arXiv:1604.06422 [astro-ph.HE]}
  \BibitemShut {NoStop}%
\bibitem [{\citenamefont {{LIGO{\ }Scientific{\
  }Collaboration}}(2015)}]{LIGOWhitePaper}%
  \BibitemOpen
  \bibfield  {author} {\bibinfo {author} {\bibnamefont {{LIGO{\ }Scientific{\
  }Collaboration}}},\ }\href
  {https://dcc.ligo.org/public/0113/T1400316/004/T1400316-v5.pdf} {\enquote
  {\bibinfo {title} {{Instrument Science White Paper, LIGO-T1400316-v4}},}\ }
  (\bibinfo {year} {2015})\BibitemShut {NoStop}%
\bibitem [{\citenamefont {Yoshino}\ and\ \citenamefont
  {Kodama}(2014)}]{Yoshino:2013ofa}%
  \BibitemOpen
  \bibfield  {author} {\bibinfo {author} {\bibfnamefont {H.}~\bibnamefont
  {Yoshino}}\ and\ \bibinfo {author} {\bibfnamefont {H.}~\bibnamefont
  {Kodama}},\ }\href {\doibase 10.1093/ptep/ptu029} {\bibfield  {journal}
  {\bibinfo  {journal} {PTEP}\ }\textbf {\bibinfo {volume} {2014}},\ \bibinfo
  {pages} {043E02} (\bibinfo {year} {2014})},\ \Eprint
  {http://arxiv.org/abs/1312.2326} {arXiv:1312.2326 [gr-qc]} \BibitemShut
  {NoStop}%
\bibitem [{\citenamefont {Aasi}\ \emph {et~al.}(2013)\citenamefont {Aasi} \emph
  {et~al.}}]{Aasi:2012fw}%
  \BibitemOpen
  \bibfield  {author} {\bibinfo {author} {\bibfnamefont {J.}~\bibnamefont
  {Aasi}} \emph {et~al.} (\bibinfo {collaboration} {VIRGO, LIGO Scientific}),\
  }\href {\doibase 10.1103/PhysRevD.87.042001} {\bibfield  {journal} {\bibinfo
  {journal} {Phys. Rev.}\ }\textbf {\bibinfo {volume} {D87}},\ \bibinfo {pages}
  {042001} (\bibinfo {year} {2013})},\ \Eprint {http://arxiv.org/abs/1207.7176}
  {arXiv:1207.7176 [gr-qc]} \BibitemShut {NoStop}%
\bibitem [{\citenamefont {Abbott}\ \emph
  {et~al.}(2016{\natexlab{b}})\citenamefont {Abbott} \emph
  {et~al.}}]{Abbott:2016udd}%
  \BibitemOpen
  \bibfield  {author} {\bibinfo {author} {\bibfnamefont {B.~P.}\ \bibnamefont
  {Abbott}} \emph {et~al.} (\bibinfo {collaboration} {Virgo, LIGO
  Scientific}),\ }\href {\doibase 10.1103/PhysRevD.94.042002} {\bibfield
  {journal} {\bibinfo  {journal} {Phys. Rev.}\ }\textbf {\bibinfo {volume}
  {D94}},\ \bibinfo {pages} {042002} (\bibinfo {year} {2016}{\natexlab{b}})},\
  \Eprint {http://arxiv.org/abs/1605.03233} {arXiv:1605.03233 [gr-qc]}
  \BibitemShut {NoStop}%
\bibitem [{\citenamefont {Abbott}\ \emph
  {et~al.}(2016{\natexlab{c}})\citenamefont {Abbott} \emph
  {et~al.}}]{TheLIGOScientific:2016uns}%
  \BibitemOpen
  \bibfield  {author} {\bibinfo {author} {\bibfnamefont {B.~P.}\ \bibnamefont
  {Abbott}} \emph {et~al.} (\bibinfo {collaboration} {Virgo, LIGO
  Scientific}),\ }\href {\doibase 10.1103/PhysRevD.94.102002} {\bibfield
  {journal} {\bibinfo  {journal} {Phys. Rev.}\ }\textbf {\bibinfo {volume}
  {D94}},\ \bibinfo {pages} {102002} (\bibinfo {year} {2016}{\natexlab{c}})},\
  \Eprint {http://arxiv.org/abs/1606.09619} {arXiv:1606.09619 [gr-qc]}
  \BibitemShut {NoStop}%
\bibitem [{\citenamefont {Okawa}\ \emph {et~al.}(2014)\citenamefont {Okawa},
  \citenamefont {Witek},\ and\ \citenamefont {Cardoso}}]{Okawa:2014nda}%
  \BibitemOpen
  \bibfield  {author} {\bibinfo {author} {\bibfnamefont {H.}~\bibnamefont
  {Okawa}}, \bibinfo {author} {\bibfnamefont {H.}~\bibnamefont {Witek}}, \ and\
  \bibinfo {author} {\bibfnamefont {V.}~\bibnamefont {Cardoso}},\ }\href
  {\doibase 10.1103/PhysRevD.89.104032} {\bibfield  {journal} {\bibinfo
  {journal} {Phys. Rev.}\ }\textbf {\bibinfo {volume} {D89}},\ \bibinfo {pages}
  {104032} (\bibinfo {year} {2014})},\ \Eprint {http://arxiv.org/abs/1401.1548}
  {arXiv:1401.1548 [gr-qc]} \BibitemShut {NoStop}%
\bibitem [{\citenamefont {Abbott}\ \emph
  {et~al.}(2016{\natexlab{d}})\citenamefont {Abbott} \emph
  {et~al.}}]{Abbott:2016nhf}%
  \BibitemOpen
  \bibfield  {author} {\bibinfo {author} {\bibfnamefont {B.~P.}\ \bibnamefont
  {Abbott}} \emph {et~al.} (\bibinfo {collaboration} {Virgo, LIGO
  Scientific}),\ }\href {\doibase 10.3847/2041-8205/833/1/L1} {\bibfield
  {journal} {\bibinfo  {journal} {Astrophys. J.}\ }\textbf {\bibinfo {volume}
  {833}},\ \bibinfo {pages} {1} (\bibinfo {year} {2016}{\natexlab{d}})},\
  \Eprint {http://arxiv.org/abs/1602.03842} {arXiv:1602.03842 [astro-ph.HE]}
  \BibitemShut {NoStop}%
\bibitem [{\citenamefont {Abbott}\ \emph
  {et~al.}(2016{\natexlab{e}})\citenamefont {Abbott} \emph
  {et~al.}}]{TheLIGOScientific:2016htt}%
  \BibitemOpen
  \bibfield  {author} {\bibinfo {author} {\bibfnamefont {B.~P.}\ \bibnamefont
  {Abbott}} \emph {et~al.} (\bibinfo {collaboration} {Virgo, LIGO
  Scientific}),\ }\href {\doibase 10.3847/2041-8205/818/2/L22} {\bibfield
  {journal} {\bibinfo  {journal} {Astrophys. J.}\ }\textbf {\bibinfo {volume}
  {818}},\ \bibinfo {pages} {L22} (\bibinfo {year} {2016}{\natexlab{e}})},\
  \Eprint {http://arxiv.org/abs/1602.03846} {arXiv:1602.03846 [astro-ph.HE]}
  \BibitemShut {NoStop}%
\bibitem [{\citenamefont {Abbott}\ \emph
  {et~al.}(2016{\natexlab{f}})\citenamefont {Abbott} \emph
  {et~al.}}]{TheLIGOScientific:2016pea}%
  \BibitemOpen
  \bibfield  {author} {\bibinfo {author} {\bibfnamefont {B.~P.}\ \bibnamefont
  {Abbott}} \emph {et~al.} (\bibinfo {collaboration} {Virgo, LIGO
  Scientific}),\ }\href {\doibase 10.1103/PhysRevX.6.041015} {\bibfield
  {journal} {\bibinfo  {journal} {Phys. Rev.}\ }\textbf {\bibinfo {volume}
  {X6}},\ \bibinfo {pages} {041015} (\bibinfo {year} {2016}{\natexlab{f}})},\
  \Eprint {http://arxiv.org/abs/1606.04856} {arXiv:1606.04856 [gr-qc]}
  \BibitemShut {NoStop}%
\bibitem [{\citenamefont {Vitale}\ \emph {et~al.}(2014)\citenamefont {Vitale},
  \citenamefont {Lynch}, \citenamefont {Veitch}, \citenamefont {Raymond},\ and\
  \citenamefont {Sturani}}]{Vitale:2014mka}%
  \BibitemOpen
  \bibfield  {author} {\bibinfo {author} {\bibfnamefont {S.}~\bibnamefont
  {Vitale}}, \bibinfo {author} {\bibfnamefont {R.}~\bibnamefont {Lynch}},
  \bibinfo {author} {\bibfnamefont {J.}~\bibnamefont {Veitch}}, \bibinfo
  {author} {\bibfnamefont {V.}~\bibnamefont {Raymond}}, \ and\ \bibinfo
  {author} {\bibfnamefont {R.}~\bibnamefont {Sturani}},\ }\href {\doibase
  10.1103/PhysRevLett.112.251101} {\bibfield  {journal} {\bibinfo  {journal}
  {Phys. Rev. Lett.}\ }\textbf {\bibinfo {volume} {112}},\ \bibinfo {pages}
  {251101} (\bibinfo {year} {2014})},\ \Eprint {http://arxiv.org/abs/1403.0129}
  {arXiv:1403.0129 [gr-qc]} \BibitemShut {NoStop}%
\bibitem [{\citenamefont {Vitale}(tion)}]{Vitale}%
  \BibitemOpen
  \bibfield  {author} {\bibinfo {author} {\bibfnamefont {S.}~\bibnamefont
  {Vitale}},\ }\href@noop {} {} (\bibinfo {year} {private
  communication})\BibitemShut {NoStop}%
\bibitem [{\citenamefont {Vitale}\ \emph {et~al.}(2016)\citenamefont {Vitale},
  \citenamefont {Lynch}, \citenamefont {Raymond}, \citenamefont {Sturani},
  \citenamefont {Veitch},\ and\ \citenamefont {Graff}}]{Vitale:2016avz}%
  \BibitemOpen
  \bibfield  {author} {\bibinfo {author} {\bibfnamefont {S.}~\bibnamefont
  {Vitale}}, \bibinfo {author} {\bibfnamefont {R.}~\bibnamefont {Lynch}},
  \bibinfo {author} {\bibfnamefont {V.}~\bibnamefont {Raymond}}, \bibinfo
  {author} {\bibfnamefont {R.}~\bibnamefont {Sturani}}, \bibinfo {author}
  {\bibfnamefont {J.}~\bibnamefont {Veitch}}, \ and\ \bibinfo {author}
  {\bibfnamefont {P.}~\bibnamefont {Graff}},\ }\href@noop {} {\  (\bibinfo
  {year} {2016})},\ \Eprint {http://arxiv.org/abs/1611.01122} {arXiv:1611.01122
  [gr-qc]} \BibitemShut {NoStop}%
\bibitem [{\citenamefont {Abbott}\ \emph
  {et~al.}(2016{\natexlab{g}})\citenamefont {Abbott} \emph
  {et~al.}}]{Abbott:1602.03846}%
  \BibitemOpen
  \bibfield  {author} {\bibinfo {author} {\bibfnamefont {B.~P.}\ \bibnamefont
  {Abbott}} \emph {et~al.} (\bibinfo {collaboration} {Virgo, LIGO
  Scientific}),\ }\href {\doibase 10.3847/2041-8205/818/2/L22} {\bibfield
  {journal} {\bibinfo  {journal} {Astrophys. J.}\ }\textbf {\bibinfo {volume}
  {818}},\ \bibinfo {pages} {L22} (\bibinfo {year} {2016}{\natexlab{g}})},\
  \Eprint {http://arxiv.org/abs/1602.03846} {arXiv:1602.03846 [astro-ph.HE]}
  \BibitemShut {NoStop}%
\bibitem [{\citenamefont {Vitale}\ and\ \citenamefont
  {Evans}(2016)}]{Vitale:2016icu}%
  \BibitemOpen
  \bibfield  {author} {\bibinfo {author} {\bibfnamefont {S.}~\bibnamefont
  {Vitale}}\ and\ \bibinfo {author} {\bibfnamefont {M.}~\bibnamefont {Evans}},\
  }\href@noop {} {\  (\bibinfo {year} {2016})},\ \Eprint
  {http://arxiv.org/abs/1610.06917} {arXiv:1610.06917 [gr-qc]} \BibitemShut
  {NoStop}%
\bibitem [{\citenamefont {Brito}\ \emph
  {et~al.}(2015{\natexlab{b}})\citenamefont {Brito}, \citenamefont {Cardoso},\
  and\ \citenamefont {Pani}}]{Brito:2014wla}%
  \BibitemOpen
  \bibfield  {author} {\bibinfo {author} {\bibfnamefont {R.}~\bibnamefont
  {Brito}}, \bibinfo {author} {\bibfnamefont {V.}~\bibnamefont {Cardoso}}, \
  and\ \bibinfo {author} {\bibfnamefont {P.}~\bibnamefont {Pani}},\ }\href
  {\doibase 10.1088/0264-9381/32/13/134001} {\bibfield  {journal} {\bibinfo
  {journal} {Class. Quant. Grav.}\ }\textbf {\bibinfo {volume} {32}},\ \bibinfo
  {pages} {134001} (\bibinfo {year} {2015}{\natexlab{b}})},\ \Eprint
  {http://arxiv.org/abs/1411.0686} {arXiv:1411.0686 [gr-qc]} \BibitemShut
  {NoStop}%
\bibitem [{\citenamefont {Belczynski}\ \emph {et~al.}(2016)\citenamefont
  {Belczynski}, \citenamefont {Repetto}, \citenamefont {Holz}, \citenamefont
  {O'Shaughnessy}, \citenamefont {Bulik}, \citenamefont {Berti}, \citenamefont
  {Fryer},\ and\ \citenamefont {Dominik}}]{Belczynski:2015tba}%
  \BibitemOpen
  \bibfield  {author} {\bibinfo {author} {\bibfnamefont {K.}~\bibnamefont
  {Belczynski}}, \bibinfo {author} {\bibfnamefont {S.}~\bibnamefont {Repetto}},
  \bibinfo {author} {\bibfnamefont {D.~E.}\ \bibnamefont {Holz}}, \bibinfo
  {author} {\bibfnamefont {R.}~\bibnamefont {O'Shaughnessy}}, \bibinfo {author}
  {\bibfnamefont {T.}~\bibnamefont {Bulik}}, \bibinfo {author} {\bibfnamefont
  {E.}~\bibnamefont {Berti}}, \bibinfo {author} {\bibfnamefont
  {C.}~\bibnamefont {Fryer}}, \ and\ \bibinfo {author} {\bibfnamefont
  {M.}~\bibnamefont {Dominik}},\ }\href {\doibase 10.3847/0004-637X/819/2/108}
  {\bibfield  {journal} {\bibinfo  {journal} {Astrophys. J.}\ }\textbf
  {\bibinfo {volume} {819}},\ \bibinfo {pages} {108} (\bibinfo {year}
  {2016})},\ \Eprint {http://arxiv.org/abs/1510.04615} {arXiv:1510.04615
  [astro-ph.HE]} \BibitemShut {NoStop}%
\bibitem [{\citenamefont {Buonanno}\ \emph {et~al.}(2008)\citenamefont
  {Buonanno}, \citenamefont {Kidder},\ and\ \citenamefont
  {Lehner}}]{Buonanno:2007sv}%
  \BibitemOpen
  \bibfield  {author} {\bibinfo {author} {\bibfnamefont {A.}~\bibnamefont
  {Buonanno}}, \bibinfo {author} {\bibfnamefont {L.~E.}\ \bibnamefont
  {Kidder}}, \ and\ \bibinfo {author} {\bibfnamefont {L.}~\bibnamefont
  {Lehner}},\ }\href {\doibase 10.1103/PhysRevD.77.026004} {\bibfield
  {journal} {\bibinfo  {journal} {Phys. Rev.}\ }\textbf {\bibinfo {volume}
  {D77}},\ \bibinfo {pages} {026004} (\bibinfo {year} {2008})},\ \Eprint
  {http://arxiv.org/abs/0709.3839} {arXiv:0709.3839 [astro-ph]} \BibitemShut
  {NoStop}%
\bibitem [{\citenamefont {Lehner}\ and\ \citenamefont
  {Pretorius}(2014)}]{Lehner:2014asa}%
  \BibitemOpen
  \bibfield  {author} {\bibinfo {author} {\bibfnamefont {L.}~\bibnamefont
  {Lehner}}\ and\ \bibinfo {author} {\bibfnamefont {F.}~\bibnamefont
  {Pretorius}},\ }\href {\doibase 10.1146/annurev-astro-081913-040031}
  {\bibfield  {journal} {\bibinfo  {journal} {Ann. Rev. Astron. Astrophys.}\
  }\textbf {\bibinfo {volume} {52}},\ \bibinfo {pages} {661} (\bibinfo {year}
  {2014})},\ \Eprint {http://arxiv.org/abs/1405.4840} {arXiv:1405.4840
  [astro-ph.HE]} \BibitemShut {NoStop}%
\bibitem [{\citenamefont {Kastaun}\ \emph {et~al.}(2013)\citenamefont
  {Kastaun}, \citenamefont {Galeazzi}, \citenamefont {Alic}, \citenamefont
  {Rezzolla},\ and\ \citenamefont {Font}}]{Kastaun:2013mv}%
  \BibitemOpen
  \bibfield  {author} {\bibinfo {author} {\bibfnamefont {W.}~\bibnamefont
  {Kastaun}}, \bibinfo {author} {\bibfnamefont {F.}~\bibnamefont {Galeazzi}},
  \bibinfo {author} {\bibfnamefont {D.}~\bibnamefont {Alic}}, \bibinfo {author}
  {\bibfnamefont {L.}~\bibnamefont {Rezzolla}}, \ and\ \bibinfo {author}
  {\bibfnamefont {J.~A.}\ \bibnamefont {Font}},\ }\href {\doibase
  10.1103/PhysRevD.88.021501} {\bibfield  {journal} {\bibinfo  {journal} {Phys.
  Rev.}\ }\textbf {\bibinfo {volume} {D88}},\ \bibinfo {pages} {021501}
  (\bibinfo {year} {2013})},\ \Eprint {http://arxiv.org/abs/1301.7348}
  {arXiv:1301.7348 [gr-qc]} \BibitemShut {NoStop}%
\end{thebibliography}%

\end{document}